\documentclass[aps,prc,amsmath,superscriptaddress,amssymb,floats,nofootinbib,twocolumn]{revtex4-1}
\usepackage[pdftex]{graphicx}
\usepackage{setspace}
\begin{document}  

\title{Polynomial fits and the proton radius puzzle}

\vspace{0.3cm}  
\author{E.~Kraus\footnote{Permanent address: Syracuse University,
    Syracuse, NY 13210, USA}}
\affiliation{Department of Physics and Astronomy, Rutgers University, Piscataway, NJ 08854, USA}
\author{K.E.~Mesick}
\affiliation{Department of Physics and Astronomy, Rutgers University, Piscataway, NJ 08854, USA}
\author{A.~White}
\affiliation{Department of Physics and Astronomy, Rutgers University, Piscataway, NJ 08854, USA}
\author{R.~Gilman} 
\affiliation{Department of Physics and Astronomy, Rutgers University, Piscataway, NJ 08854, USA}
\author{S.~Strauch}
\affiliation{Department of Physics and Astronomy, University of South Carolina, Columbia, SC 29208, USA}

\begin{abstract}
The Proton Radius Puzzle refers to the $\approx$7$\sigma$ discrepancy that exists
between the proton charge radius determined from muonic hydrogen and that determined from
electronic hydrogen spectroscopy and electron-proton scattering.
One possible partial resolution to the puzzle includes errors in the
extraction of the proton radius from $ep$ elastic scattering data.
This possibility is made plausible by certain fits which extract
a smaller proton radius from the scattering data consistent with that determined from
muonic hydrogen.
The reliability of some of these fits that yield a smaller proton radius was studied.
We found that fits of form factor data with a truncated polynomial fit are unreliable
and systematically give values for the proton radius that are too small.  
Additionally, a polynomial fit with a $\chi^2_{reduced} \approx 1$
is not a sufficient indication for a reliable result.
\end{abstract}

\maketitle


\section{Physics Motivation}


The {\em Proton Radius Puzzle} pertains to the disagreement between the 
proton charge radius determined from muonic hydrogen and from
electron-proton systems: atomic hydrogen and $ep$ elastic scattering.
The muonic hydrogen result~\cite{Pohl:2010zza,Antognini25012013} of
$r_p = 0.84087 \pm 0.00039$~fm is about 13 times more precise and
$\approx$7$\sigma$ different than the recent CODATA 2010~\cite{Mohr:2010} result of
$r_p = 0.8775 \pm 0.0051$~fm.
The CODATA analysis includes atomic hydrogen and the precise
cross section measurements of Bernauer {\it et  al.}~\cite{Bernauer:2010wm,Bernauer:2014},
which give $r_p = 0.879 \pm 0.008$~fm, but not the more recent
confirmation of Zhan {\it et al.}~\cite{Zhan:2011ji} which yields
$r_p = 0.875 \pm 0.010$~fm.
For a recent review, see~\cite{Pohlreview:2012}.

Many possible explanations of the Proton Radius Puzzle have been ruled
out. There are, for example, no known issues with the atomic theory,
or with the muonic hydrogen experiment.
It appears that the most likely explanations are 
novel physics beyond the Standard Model that differentiates $\mu p$ and $ep$ interactions,
novel two-photon exchange effects that differentiate $\mu p$ and $ep$ interactions,
and errors in the $ep$ experiments. 
It is therefore important to examine possible issues in the $ep$ experiments
before concluding that interesting physics is required.

\begin{figure}[h]
\centerline{\includegraphics[width=2.8in]{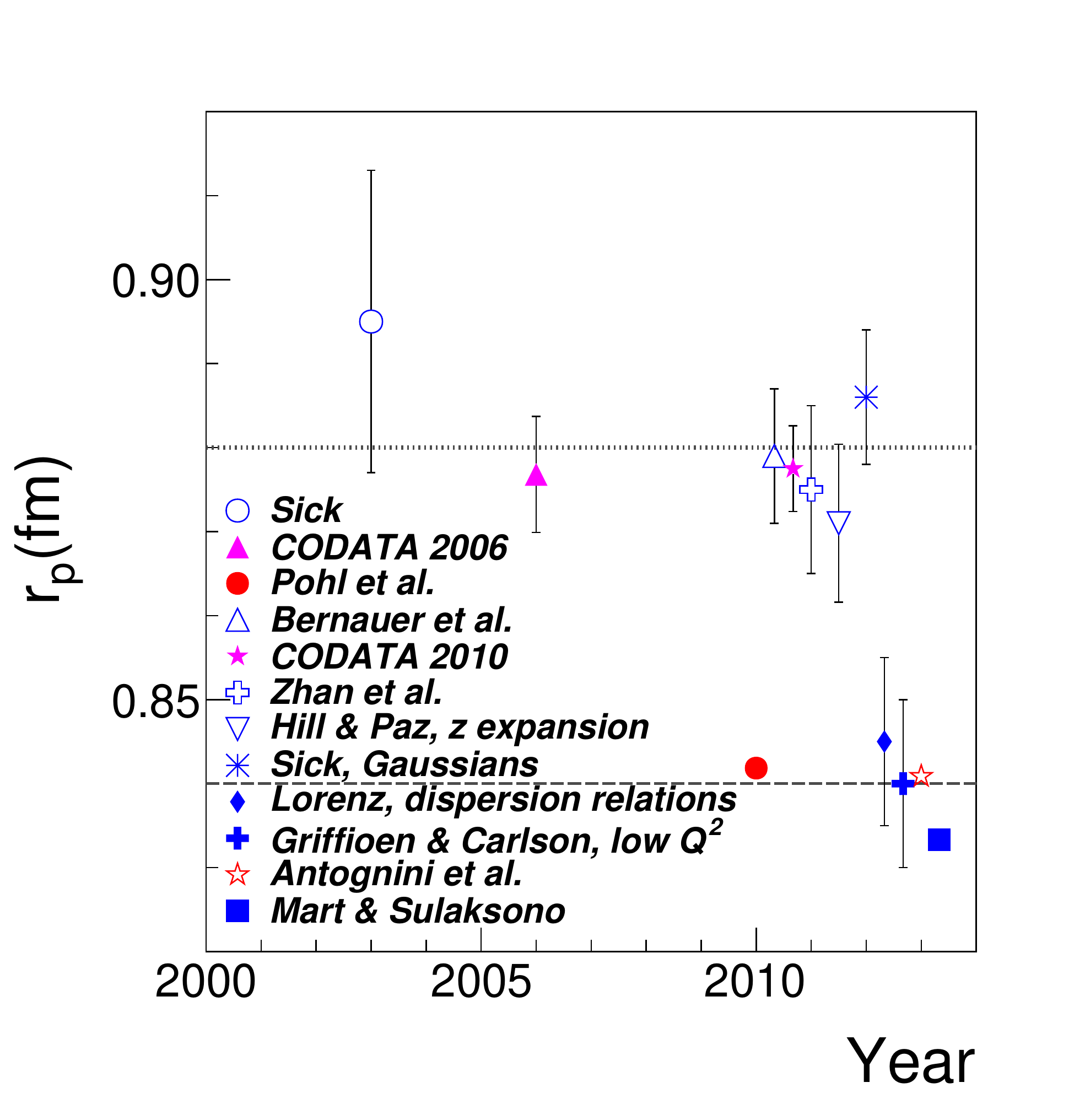}}
\caption{(Color online) 
A summary of some recent proton charge radius
  determinations: Sick~\cite{Sick:2003gm}, 
CODATA 2006~\cite{Mohr:2008fa}, 
Pohl {\it et al.}~\cite{Pohl:2010zza},
Bernauer {\it et al.}~\cite{Bernauer:2010wm,Bernauer:2014}, 
CODATA 2010~\cite{Mohr:2010}, 
Zhan {\it et al.}~\cite{Zhan:2011ji},
Hill \& Paz~\cite{Hill:2010yb},
Sick Gaussians~\cite{Sick:2011zz,Sick:2012zz},
Lorenz {\it et al.}~\cite{Lorenz:2012tm},
Griffioen \& Carlson~\cite{GriffioenCarlson},
Antognini {\it et al.}~\cite{Antognini25012013}
and
Mart \& Sulaksono~\cite{PhysRevC.87.025807}.
The dashed and dotted lines are drawn at 0.88 fm and 0.84 fm, respectively, for reference.
}
\label{fig:radius}
\end{figure}

While the extracted radius values given above have been confirmed by
some analyses, other analyses of $ep$ scattering data 
give a smaller radius consistent with
the muonic hydrogen result. 
Examples of confirming analyses include
the $z$ expansion of~\cite{Hill:2010yb}
($r_p$ = 0.871 fm $\pm$ 0.009 fm $\pm$ 0.002 fm $\pm$ 0.002 fm),
and
the sum-of-Gaussians fit of~\cite{Sick:2011zz,Sick:2012zz,PhysRevC.89.012201}
($r_p$ = 0.886 fm $\pm$ 0.008 fm).
However, three recent analyses give smaller radii, consistent with the muonic
hydrogen result.
Griffioen and Carlson~\cite{GriffioenCarlson} observed that a truncated linear polynomial fit of the low $Q^2$ Bernauer data yields $r_p$ $\approx$ 
0.84 fm, with good $\chi^2$.
The dispersion relation analysis of Lorenz {\it et al.}~\cite{Lorenz:2012tm} yields
$r_p$ = 0.84 $\pm$ 0.01 fm with a large $\chi^2_{reduced}$ $\approx$
2.2, in a simultaneous fit of proton and neutron data.
The fluctuating radius fit of \cite{PhysRevC.87.025807} yields
$r_p$ = 0.8333 $\pm$ 0.0004 fm with $\chi^2_{reduced}$ $\approx$ 4 -- but note
the criticism of \cite{downiecomment}.
A summary of some recent proton radius determinations can be seen in Fig.~\ref{fig:radius}.
The variation in the radius determined from scattering experiments
calls into question the reliability of the proton radius determination
from scattering experiments.

In this paper we study the reliability of proton radius
determinations from the $ep$ elastic scattering experiments.
We note that there are a number of issues in extracting a radius
from the experimental data, as discussed in \cite{Pohlreview:2012} and
\cite{PhysRevC.89.012201}.
In particular, we look at the radius extraction through the
Taylor series expansion of the proton electric form factor:
$G^p_E(Q^2) = 1 - Q^2 r_p^2/6 + Q^4 r_p^4/120 + \ldots$ such that
$r_p^2 = -6 d G^p_E(Q^2) / dQ^2|_{Q^2=0}$.  
We use a polynomial fit\footnote{Note that in similar analyses, the polynomial fit is commonly called the Taylor series expansion.} that has the same functional form as a truncated Taylor series expansion, and note
that a polynomial fit
exhibits unphysical behavior in extrapolations to large $Q^2$, as it necessarily 
diverges to infinity, and this might also affect a radius determination.

The basic result of this paper -- that radius extractions
with polynomial fits cannot be trusted to be reliable --
has already been argued by Sick \cite{Sick:2003gm},
who claimed that higher-order terms in the expansion prevent a
precise determination of the proton radius for any $Q^2$
region. Determining the $Q^2$ term precisely
requires a larger $Q^2$ range to determine the $Q^4$ term precisely,
which requires an even larger $Q^2$ range to determine the $Q^6$ term
precisely, \textit{etc}.
The inefficiency and inconsistency of the truncated polynomial fit 
has also been demonstrated in unpublished numerical work by Distler \cite{distlerpc}.
Lastly, Borisyuk \cite{Borisyuk} has argued that there is a systematic error related to the
deviation of a fitted radius and the true radius due to the inadequacy of the form factor
parameterizations in describing the true form factor.
In this paper, we find with the polynomial fits an offset between the real radius
and the radius extracted with a fit, that results from truncating the power series expansion to fit
a finite range of data. This is a systematic error that we call the {\em truncation offset}.

\section{Method}

\begin{figure}[tbh]
\centerline{\includegraphics[width=2.8in]{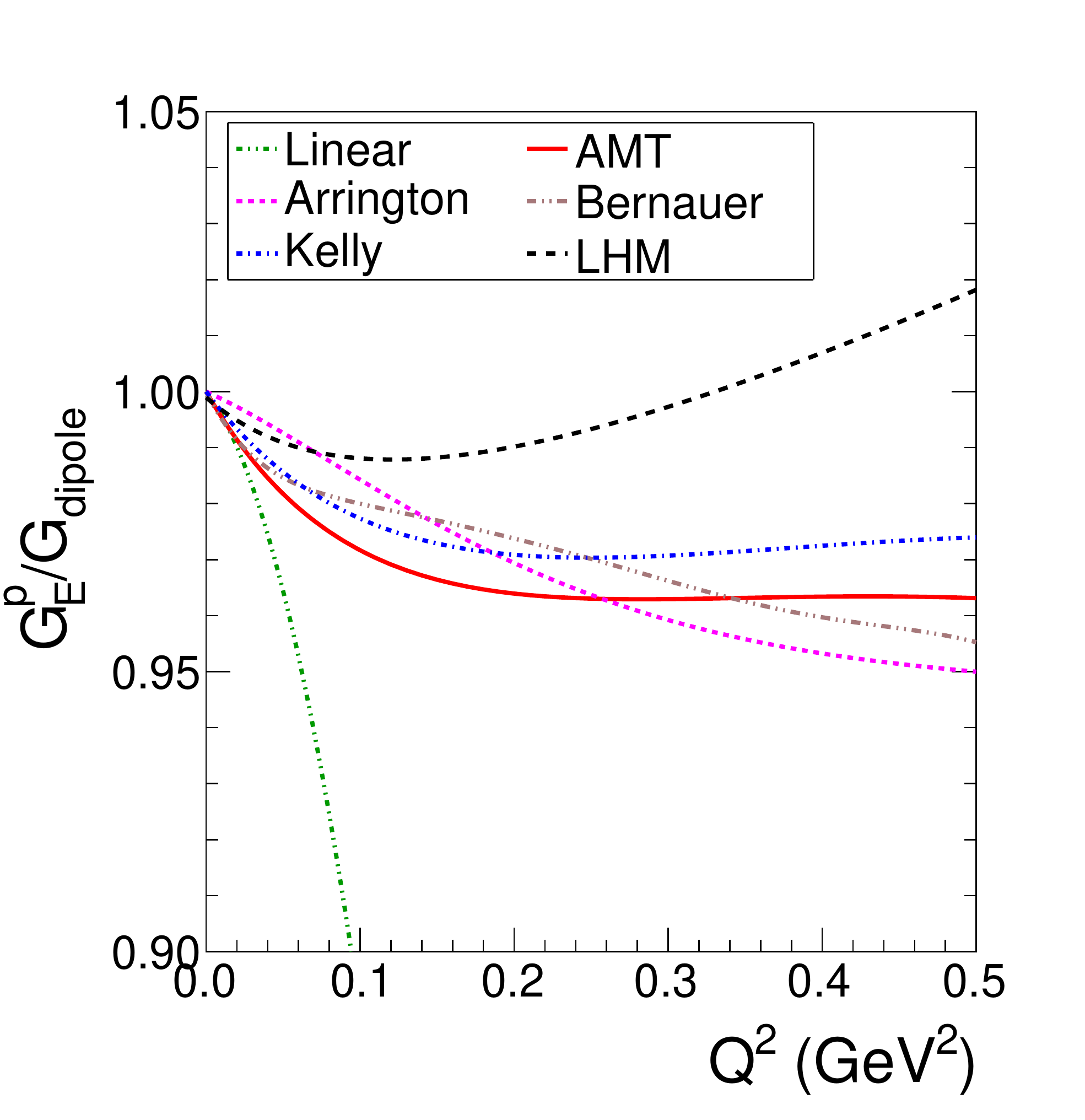}}
\caption{(Color online) Parameterizations of the proton electric form factor
used to generate pseudodata, relative to the dipole form factor,
$G_{dipole} = (1+Q^2/0.71\,{\rm GeV}^2)^{-2}$.
}
\label{fig:pfuncs}
\end{figure}

The most precise $ep$ elastic scattering data come from
Bernauer {\it et  al.} \cite{Bernauer:2010wm,Bernauer:2014}, but for our purposes
it is more useful to generate pseudodata for $G_E^p$ from a
parameterization with a known radius.
To get data similar in shape to the actual proton form factor,
and to study how sensitive the result is to the input, we generate 
the pseudodata from six parameterizations
of the proton form factor data:
\begin{itemize}
\item the Arrington, Melnitchouk, Tjon (AMT) fit \cite{PhysRevC.76.035205},
 a Pad\'{e} parameterization with $r_p \approx 0.878$~fm,
\item the Arrington fit \cite{PhysRevC.69.022201}, an inverse
  polynomial parameterization with $r_p \approx 0.829$~fm,
\item the Bernauer $n = 10$ polynomial fit \cite{Bernauer:thesis}, with
$r_p \approx 0.887$~fm,
\item the standard dipole fit, with $r_p \approx 0.811$~fm,
\item the Kelly fit \cite{Kelly:2004hm}, a Pad\'{e} parameterization
  with $r_p \approx 0.863$~fm, and
\item the Lorenz, Hammer, and Meissner (LHM) fit \cite{Lorenz:2012tm}, which
 combines dispersion relations with a vector meson dominance
 parameterization with $r_p \approx 0.84$~fm.
\end{itemize}
In addition, the numerical procedures were confirmed by generating
pseudodata from a linear function with $r_p$ = 0.86~fm.
Figure~\ref{fig:pfuncs} compares the parameterizations listed above.

For each form factor parameterization, we generate pseudodata points with 0.2\%
uncertainties (corresponding to 0.4\% cross section uncertainties) 
spaced every 0.001 GeV$^2$ in $Q^2$ from 
$Q^2_{min}$ = 0.004 GeV$^2$ to a variable $Q^2_{max}$. 
These pseudodata have roughly the uncertainties and data-point 
density of the Bernauer data, but reflect a known radius.
We fit the data with polynomials in $Q^2$:
\begin{equation}\label{eq:poly}
a_0 \left[1 + \sum\limits_{i=1}^n a_i (Q^2)^i\right]~,
\end{equation}
with $n$ = 1, 2, 3, and 4, and where $a_0$ was statistically consistent with unity
and $a_1 \propto r^2$.
For each parameterization, polynomial order, and $Q^2_{max}$, the
pseudodata generation and fitting 
is repeated 5000 times to generate distributions of $r^2$,
$\sigma (r^2)$, and $\chi^2$.
From these distributions we extract the proton charge radius and
its uncertainty, and the mean $\chi^2$. 
Numerical work was done using CERN MINUIT and ROOT.

\section{Results}

\begin{figure}[tbh]
\centerline{\includegraphics[width=2.7in]{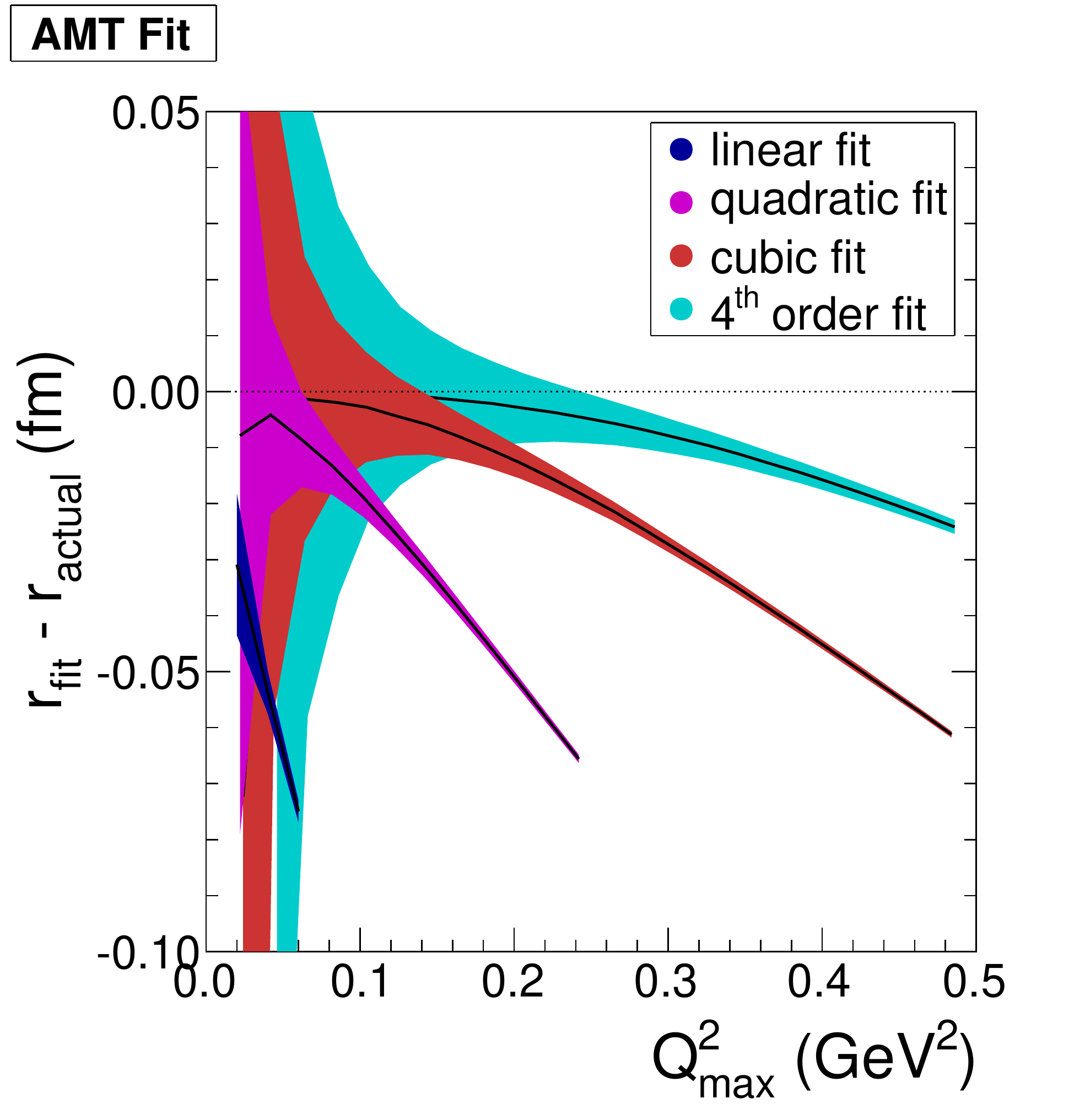}}
\caption{(Color online) The truncation offset versus $Q^2_{max}$ for the AMT
parameterization. The lines indicate the size of the truncation offset and
the bands the r.m.s.\ width of the radius distribution from the 5000 fits.
}
\label{fig:fitresults}
\end{figure}

\begin{figure}[tbh]
\centerline{\includegraphics[width=2.7in]{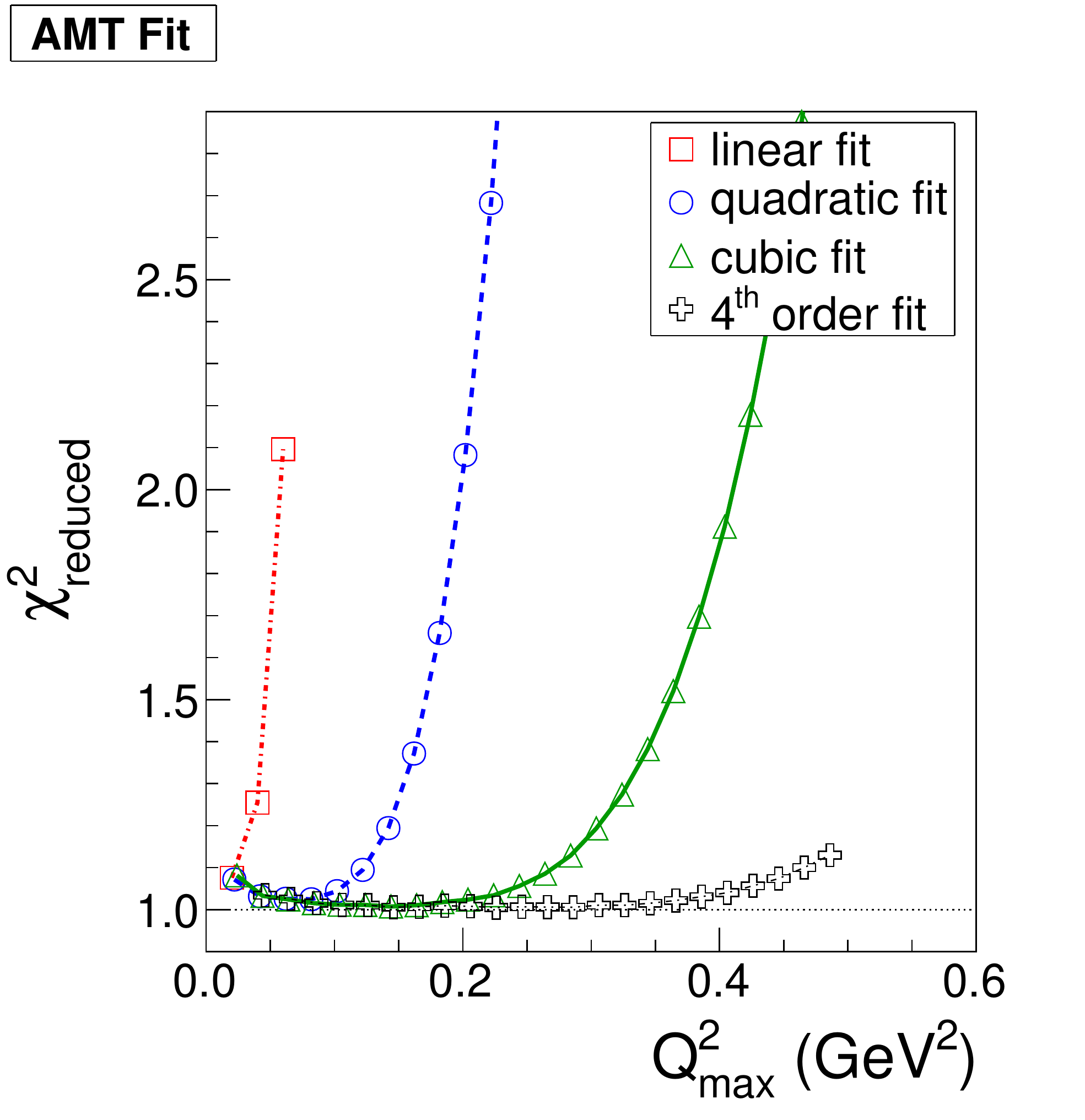}}
\caption{(Color online) Reduced $\chi^2$ for the fits of the pseudodata generated 
from the AMT form factor parameterization versus $Q^2_{max}$.
}
\label{fig:fitresults2}
\end{figure}

\begin{figure}[h]
\centerline{\includegraphics[width=2.7in]{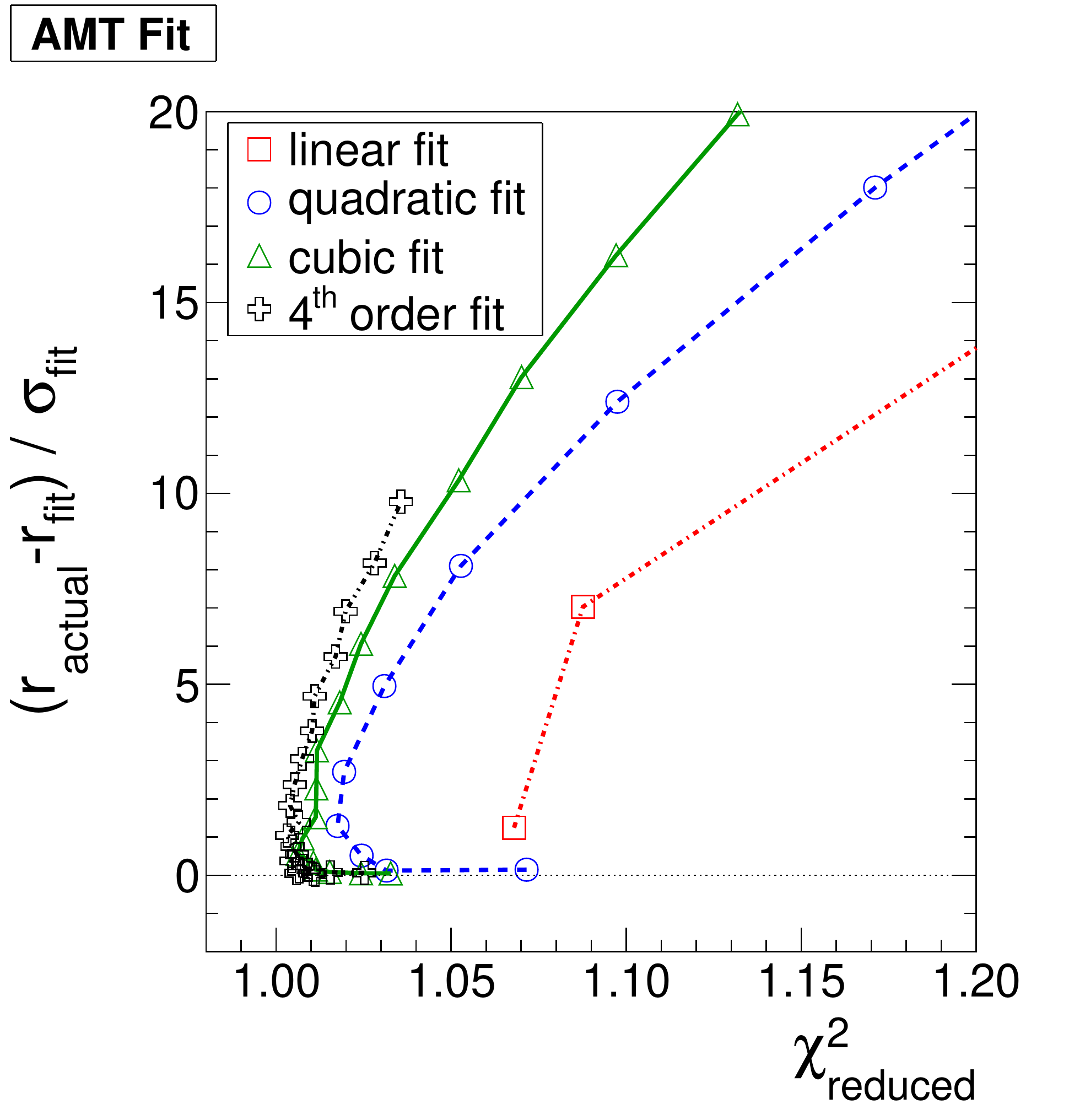}}
\caption{(Color online) The truncation offset divided by the fit
uncertainty as a function of the fit $\chi^2_{reduced}$, for the
AMT parameterization.
}
\label{fig:fitresults6}
\end{figure}

We find the results from all six form factor parameterizations are
qualitatively similar.
The AMT pseudodata fits are representative of the typical behavior
and are shown here.
Figure~\ref{fig:fitresults} shows the truncation offset versus $Q^2_{max}$.
The lines shown indicate the truncation offset, while the width of the
bands indicates 
the r.m.s.\ width of the distribution of proton radii
from the 5000 fits done, corresponding to the statistical 
uncertainty of the radius extraction in the fit.
Figure~\ref{fig:fitresults2} shows how $\chi^2_{reduced}$ varies
with $Q^2_{max}$.  Lastly, 
Fig.~\ref{fig:fitresults6} shows the truncation error in units of the fit uncertainty versus $\chi^2_{reduced}$.
In all plots, the four series of fits shown correspond to the polynomials of order
1 to 4 as defined in Eq.~(\ref{eq:poly}).


Some observations related to these figures include:
\begin{itemize}
\item Fit uncertainties decrease with increasing $Q^2_{max}$ due
  to the greater number of data points and the greater ``lever arm''
  of the data.

\item Low $Q^2$ data with the uncertainties and data-point density we have assumed
do not by themselves determine a precise radius.

\item As there is more curvature in the generating functions than in
  the fit functions, the truncation offset generally grows
  with $Q^2_{max}$, but decreases with increasing order of the fit.

\item The nature of the curvature in the proton electric form factor
 is such that the truncation offset using these parameterizations
 almost always leads to a fit radius that
 is smaller than the ``real'' radius.

\item Comparing Figs.~\ref{fig:fitresults} and \ref{fig:fitresults2} 
shows that {\em $\chi^2_{reduced}$ is not a reliable guide to the quality of the
radius extracted}.  There can already be a significant truncation offset before the  $\chi^2_{reduced}$ is obviously far from unity.
For example, in the cubic AMT fit with $Q^2_{max}$ = 0.24~GeV$^2$,
$\chi^2_{reduced}$ = 1.018, but the extracted radius of 0.859 $\pm$ 0.002~fm
differs from the 0.878 fm radius of the AMT fit by 0.019 fm, about half
of the proton-radius-puzzle discrepancy, and about 10 times the fit uncertainty.

\item The above point is also demonstrated in Fig.~\ref{fig:fitresults6}, which shows
that a small $\chi^2_{reduced}$ does not guarantee an accurate determination
of the radius; even with a small truncation error, the truncation offset of the
fit is several times the fit uncertainty.

\item Even fits with $\chi^2_{reduced}$ $<$ 1.1 can result 
in a truncation offset equal to the difference between the
 $ep$ and $\mu p$ proton radius determinations,
 $\Delta r$ $\approx$ 0.037~fm.
 
\item One can find combinations of $Q^2_{max}$ and fit order for 
which there is no significant truncation offset and good
statistical precision on the extracted radius (as done in \cite{Borisyuk}, but with
a different fitting parameterization). However, the
combinations vary with form factor parameterization, and it is
problematic in practice to ensure that a radius extracted with
a polynomial fit from actual data is reliable.

\end{itemize}

To summarize, a proton radius determination through a polynomial fit analysis is suspect.
It is believed that other fit functions, such as the inverse polynomial
or z-expansion have smaller, but still significant, truncation offsets \cite{arringtonpc}.

\begin{figure}[tbh]
\centerline{\includegraphics[width=2.7in]{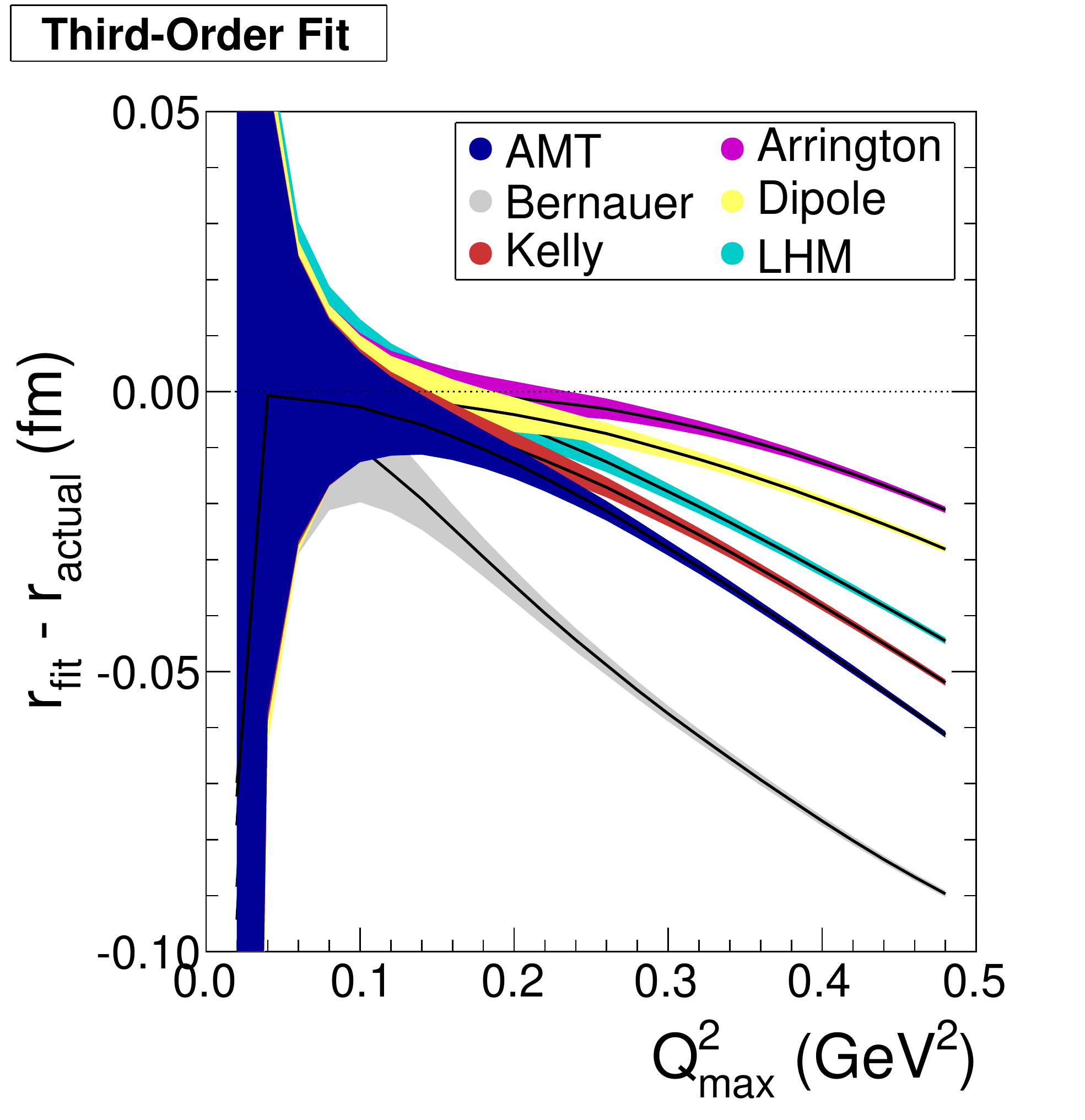}}
\caption{(Color online) The truncation offset versus $Q^2_{max}$ for the third-order fits of the pseudodata generated 
from the different form factor parameterizations.
}
\label{fig:fitresults3}
\end{figure}

\begin{figure}[tbh]
\centerline{\includegraphics[width=2.7in]{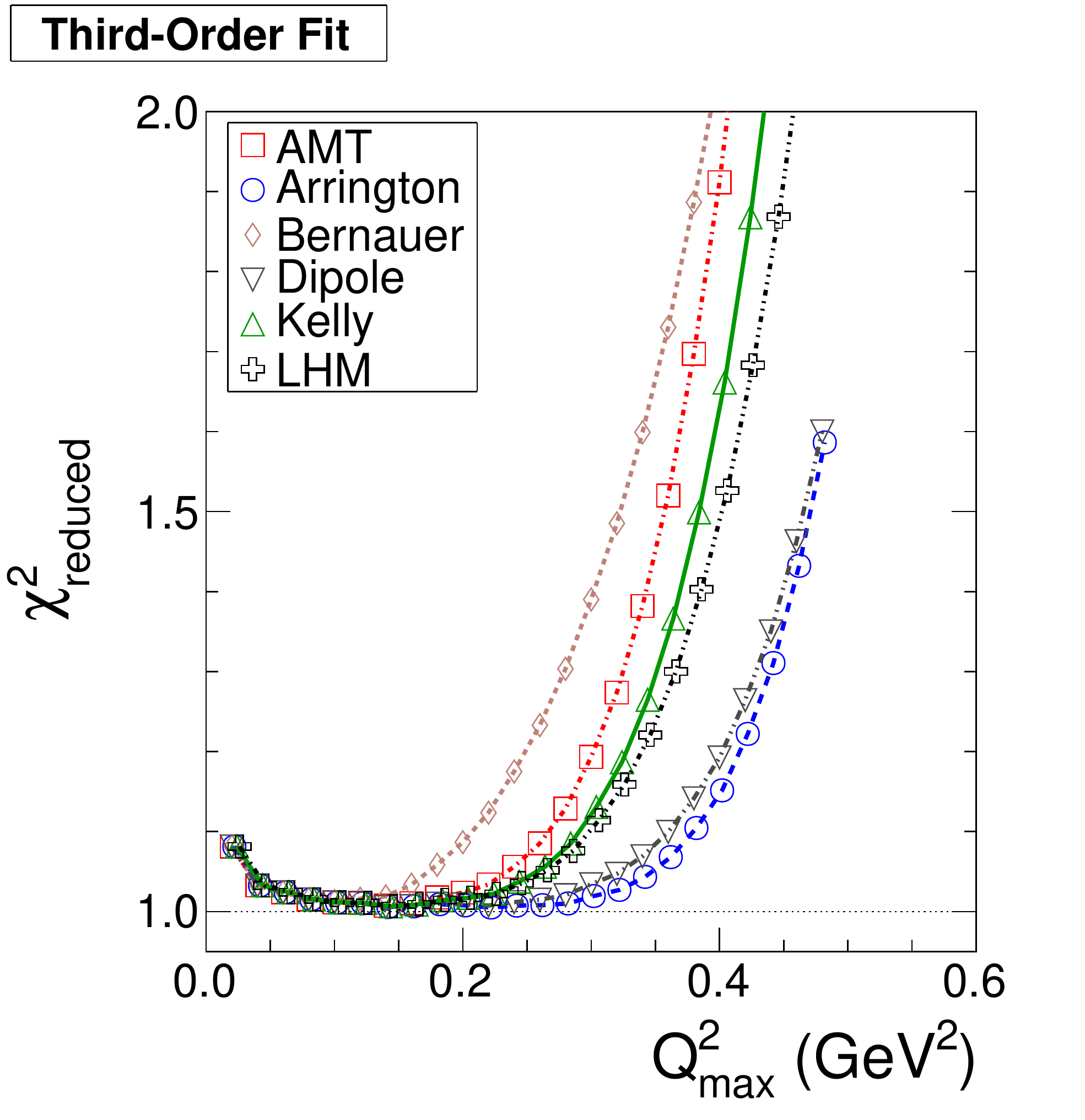}}
\caption{(Color online) Reduced $\chi^2$ versus $Q^2_{max}$ for the third-order fits of the pseudodata generated 
from the different form factor parameterizations.
}
\label{fig:fitresults4}
\end{figure}

\begin{figure}[h]
\centerline{\includegraphics[width=2.7in]{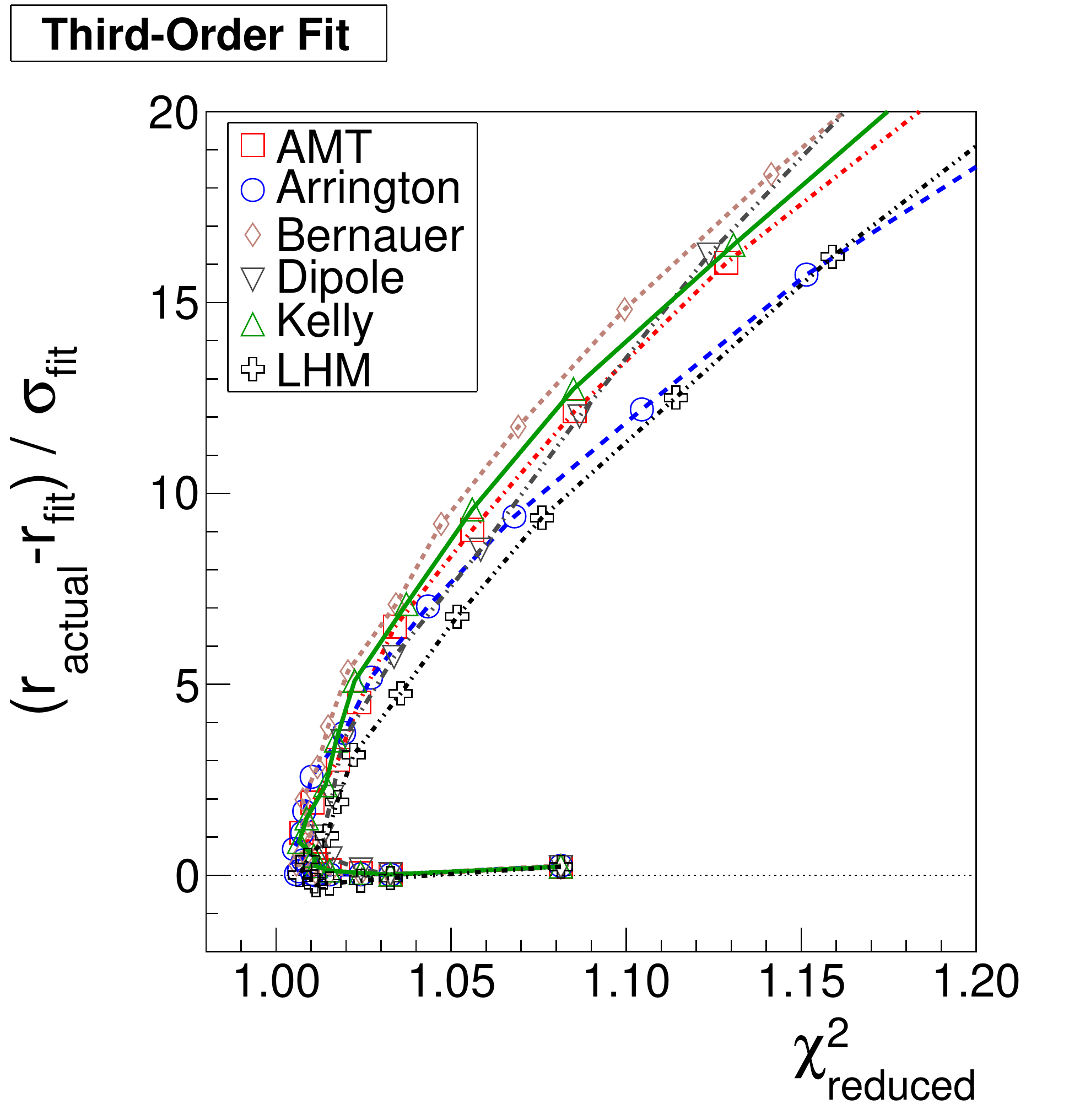}}
\caption{(Color online) The truncation offset divided by the fit
uncertainty as a function of the fit $\chi^2_{reduced}$, for the
six different parameterizations.
}
\label{fig:fitresults8}
\end{figure}

As mentioned, six different form factor parameterizations were studied.
Figures~\ref{fig:fitresults3} -- \ref{fig:fitresults8}
compare the third-order cubic fit results for all the form factor parameterizations.
Fits to a form factor following the Arrington parameterization give
the smallest truncation offset and is the least sensitive to fit order, while fits to a form factor following the
Bernauer polynomial parameterization result in the largest truncation offset and sensitivity to fit order.


\begin{figure}[tbh]
\centerline{\includegraphics[width=2.7in]{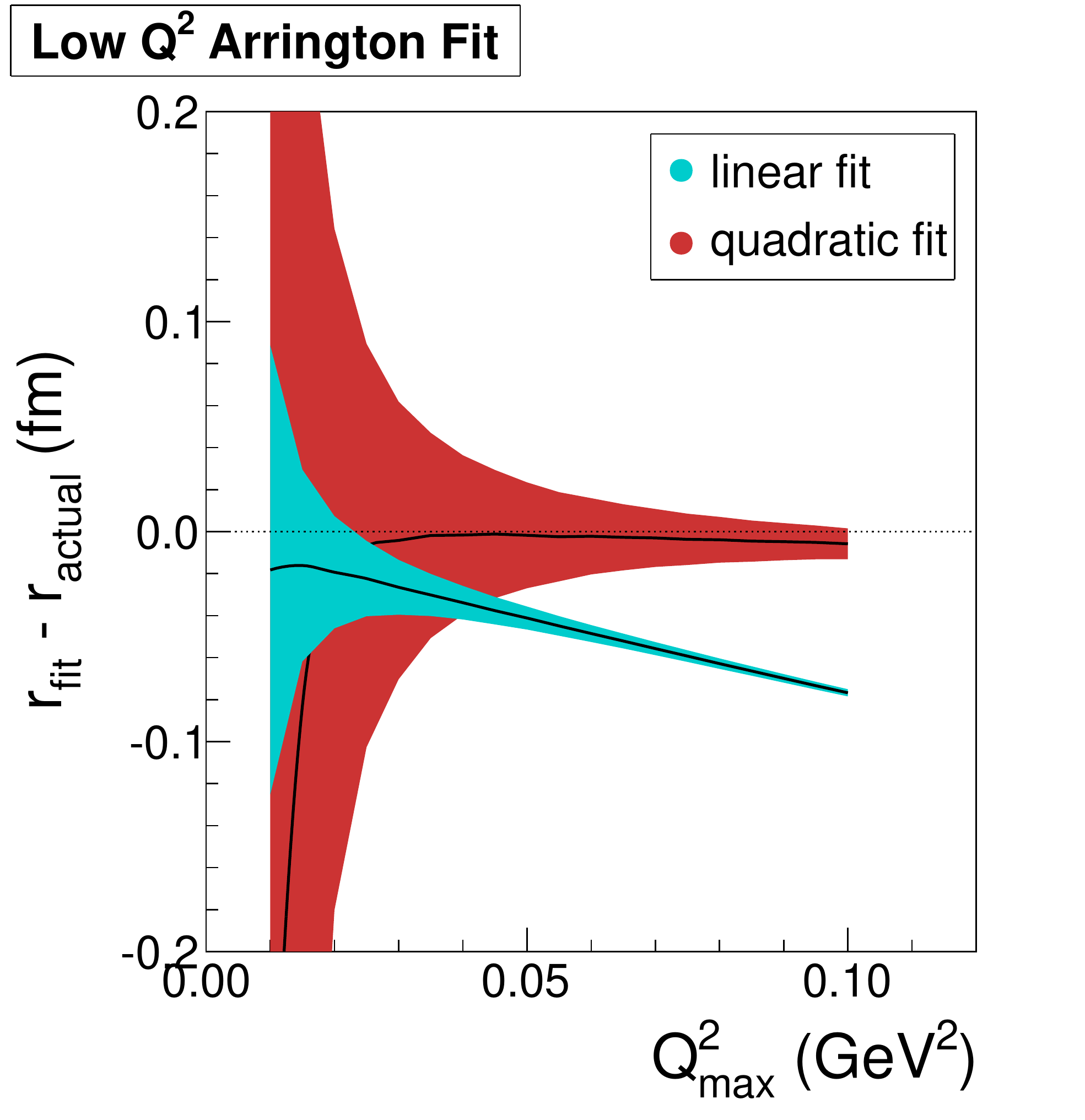}}
\caption{(Color online) The truncation offset versus $Q^2_{max}$ for
liner and quadratic fits in the low $Q^2$ region.}
\label{fig:lowq2}
\end{figure}

\begin{figure}[tbh]
\centerline{\includegraphics[width=2.7in]{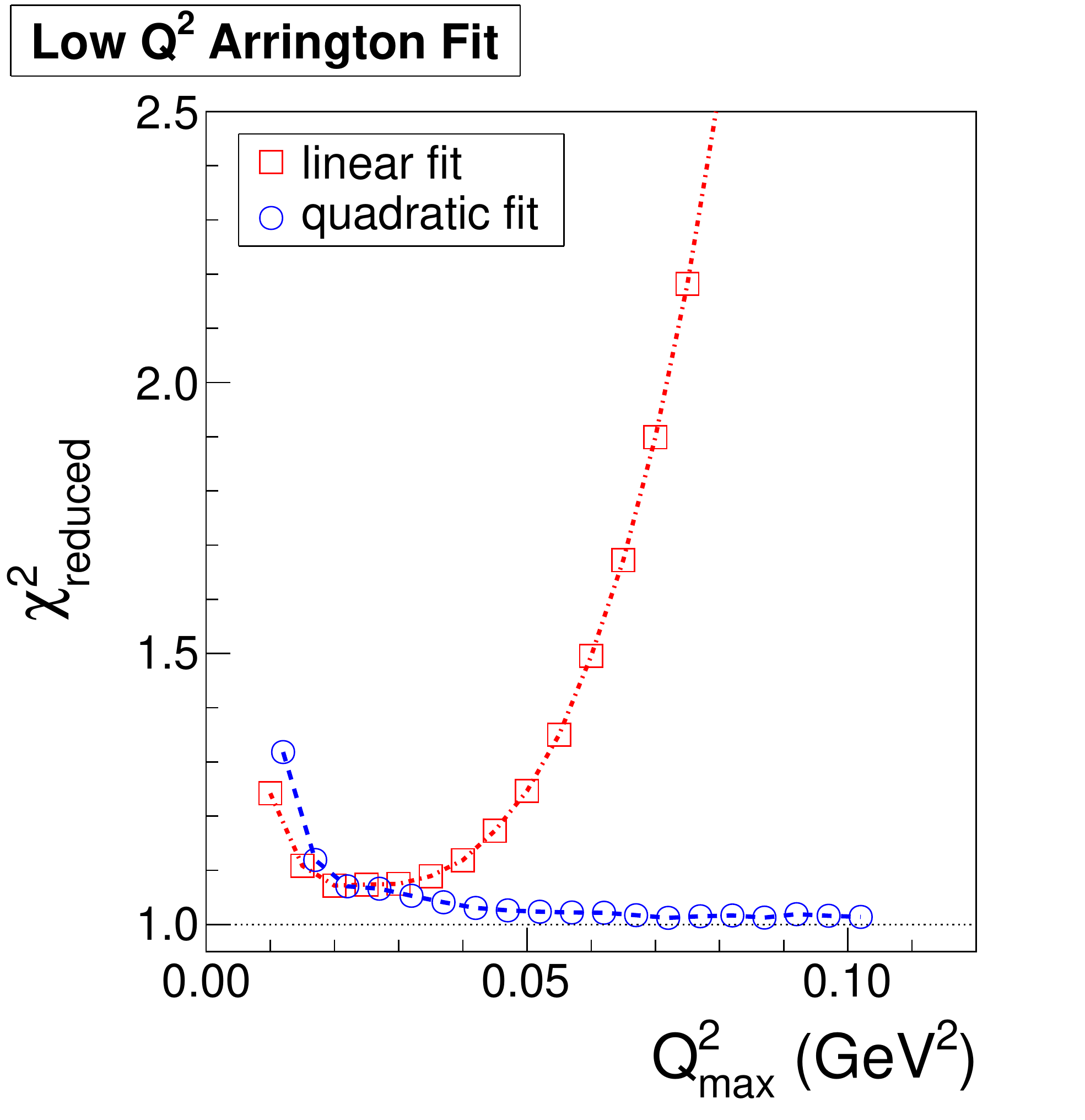}}
\caption{(Color online) Reduced $\chi^2$ versus $Q^2_{max}$ for linear 
and quadratic fits in the low $Q^2$ region.}
\label{fig:lowq22}
\end{figure}

Also of interest is how the truncation offset might affect upcoming
experiments should a truncated polynomial fit be used,
in particular low $Q^2$ measurements of the proton radius.
In this case the region of interest is $Q^2_{max} < 0.1$ GeV$^2$.  The 
results fitting from $Q^2_{min} = 0.004$~GeV$^2$ up to $Q^2_{max} = 0.01 - 0.1$~GeV$^2$ are shown for fits of 
order 1 and 2 in
Figs.~\ref{fig:lowq2} and \ref{fig:lowq22} for the truncation offset and the $\chi^2_{reduced}$, respectively, 
using the Arrington parameterization.  
Ultimately, the statistical
fit uncertainty on extracting the radius depends on the final uncertainties the future
experiments achieve, however, the truncation error depends on the $Q^2$ range.


One upcoming experiment is Jefferson Lab E12-11-106 
\cite{gasparian2011}, which plans to measure elastic $ep$ 
scattering in the range $Q^2\approx$10$^{-4}$ -- 0.02 GeV$^2$.
We simulate the experiment using 12 data points at the $Q^2$
values shown in Fig.~30 of the proposal -- note that other
estimates in the proposal re-bin the data into more points.
Under these assumptions, a linear fit to pseudodata yields a 
truncation offset ranging from 0.016~fm for the Arrington parameterization 
to 0.025~fm when the AMT parameterization is used.  The $\chi^2_{reduced}$ for both
fit examples is $\approx$1.1.  A higher-order quadratic fit reduces the truncation
offset by an order of magnitude, however results in a statistical fit uncertainty
of 0.05~fm assuming 0.4\% point-to-point cross section uncertainties.
This demonstrates that a truncated polynomial fit of the E12-11-106 data alone
is highly suspect as a technique to determine an accurate radius.

A second upcoming experiment is the MUSE measurement of
$\mu^{\pm}p$ and $e^{\pm}p$ scattering at the Paul Scherrer Institute \cite{Gilman:2013eiv}.
This experiment will have 6 independent datasets (3 different beam momentum, 2 polarities) for each particle type,
covering a $Q^2$ range of 0.0025 -- 0.0775 GeV$^2$.
MUSE will make a relative comparison of the $ep$ and $\mu p$ elastic scattering cross sections
and form factors, largely canceling several
systematic uncertainties including the truncation offset in the radius extraction.
Doing so will allow 
for a $\approx$0.01~fm measurement of the difference between the proton charge 
radius as measured by electrons versus muons.

In summary, Sick \cite{Sick:2003gm} and Distler \cite{distlerpc} have indicated that a precise proton radius could
not reliably be extracted using a polynomial fit of the form
factor. Using six form factor parameterizations for which the
radius is known, we have confirmed that this is the case. In particular, we
have shown that the condition $\chi^2_{reduced}$ $\approx$ 1 is not
sufficient for the extracted radius to be reliable.
Due to the higher-order terms in the polynomial fit, 
even an apparently good fit of the data can have a significant offset from the real radius.
This truncation offset increases with fitting a wider range of data, but decreases
with fitting with a higher-order expansion.  Even for a fit with a 
small truncation offset, the offset can be large relative to the fit uncertainty.
Finally, we have also observed that for the six form factor parameterizations used
to generate pseudodata, the truncation offset generally results in an extracted radius 
that is smaller than the true radius.  


This work was supported by the US National Science Foundation
through grants PHY 1263280 and PHY 1306126 to Rutgers University,
and through grant PHY 1205782 to the University of South Carolina.
We thank Andrew Jiang for his assistance with the calculation in this paper.


\bibliography{truncationpaper_submitv4}

\end{document}